\documentclass[review]{elsarticle}

\usepackage{lineno,hyperref}
\modulolinenumbers[5]

\journal{Journal of \LaTeX\ Templates}

\begin{document}
\begin{frontmatter}


\title{Bloch states in light transport through a perforated
metal}

\author[mymainaddress,mysecondaryaddress]{Zhyrair Gevorkian}
\address[mymainaddress]{Yerevan Physics Institute,Alikhanian Brothers St. 2,0036 Yerevan, Armenia.}
\address[mysecondaryaddress]{Institute of Radiophysics and Electronics,Ashtarak-2,0203,Armenia.}

\author{Vladimir Gasparian}
\address{California State University, Bakersfield, California 93311-1022, USA}
\author{Emilio Cuevas}
\address{Departamento de F{\'\i}sica, Universidad de Murcia, E-30071 Murcia, Spain}


\cortext[mycorrespondingauthor]{Zhyrair Gevorkian, gevork@yerphi.am}







\begin{abstract}
 Light transport in a metal with two-dimensional hole arrays is considered.
Analytical expression for a transmission coefficient in periodic, isolated and disordered cases are derived, assuming the existence of transverse waveguide modes tunneling in two-dimensional plane perpendicular
to traveling direction of light. The one dimensional case of periodic holes, due to its simplicity, is investigated in detail. In the dilute metal regime, when metal fraction is small, our numerical study of the transmission coefficient of central diffracted wave indicates the existence of a minimum which is completely independent of an incident wavelength. Further increasing of metal fraction leads to the unusual monotonic increasing of central diffracted wave transmission. The role of the surface plasmons is discussed.
\end{abstract}
\begin{keyword}
\texttt{periodical hole arrays, Bloch states}
\end{keyword}
\end{frontmatter}


\section{Introduction}
Since its discovery \cite{EOT} the extraordinary optical transmittance (EOT) attracts great interest. This interest is largely motivated by a recent progress in nanotechnology which allows to get for EOT a possible applications in different optical devices. Many experimental and theoretical papers devoted to a study of the EOT (for a review, see Ref.\cite{rev}). The phenomenon of EOT seemed to be well understood \cite{Mor01,mictheory} with the involving of the surface plasmons and Bloch states. The former appeared on the interface between metal and dielectric, while the Bloch states originating by the periodicity of hole arrays.
However, so far in theoretical understanding of EOT much attention was paid to plasmon aspect of the problem \cite{nphys364,nphys372} and less attention to the Bloch states and periodicity. The Bloch states of plasmons on the periodically perforated metal surface were studied in \cite{darman}. It has shown, that EOT phenomenon can be explained using the mechanism of plasmons' vertical tunneling from one metal surface to another one, where plasmons are eventually converted to free photon states.

In the present paper, without pretending to a complete mathematical
description the theory of EOT, we develop a different approach to
study the behavior of electromagnetic waves in a periodically performed metal system taking into account the existence of transverse waveguide modes tunneling in two-dimensional surface perpendicular
to traveling direction of light. We have concentrated our attention on the role of Bloch states and show that this mechanism with a transverse waveguide modes tunneling leads to peculiarities in light transport. Particularly, in the dilute case when metal fraction is small, we have found an independence of transmission coefficient of central diffracted wave on the incident wavelength and its unusual increase with increasing of a metal fraction. We explicitly derived an analytical expression for transmission coefficient in disordered hole arrays case. We found that without Bloch states the disordered hole arrays lead to broadening of spectral shape in accordance with the Refs. \cite{AuB09,PrGE12}.

\section{Formulation of the problem}
Let us consider a metallic film with periodic array of two dimensional holes (see Fig.1).
\begin{figure}
\vspace{-3cm}
 \begin{center}
\includegraphics[width=8.0cm]{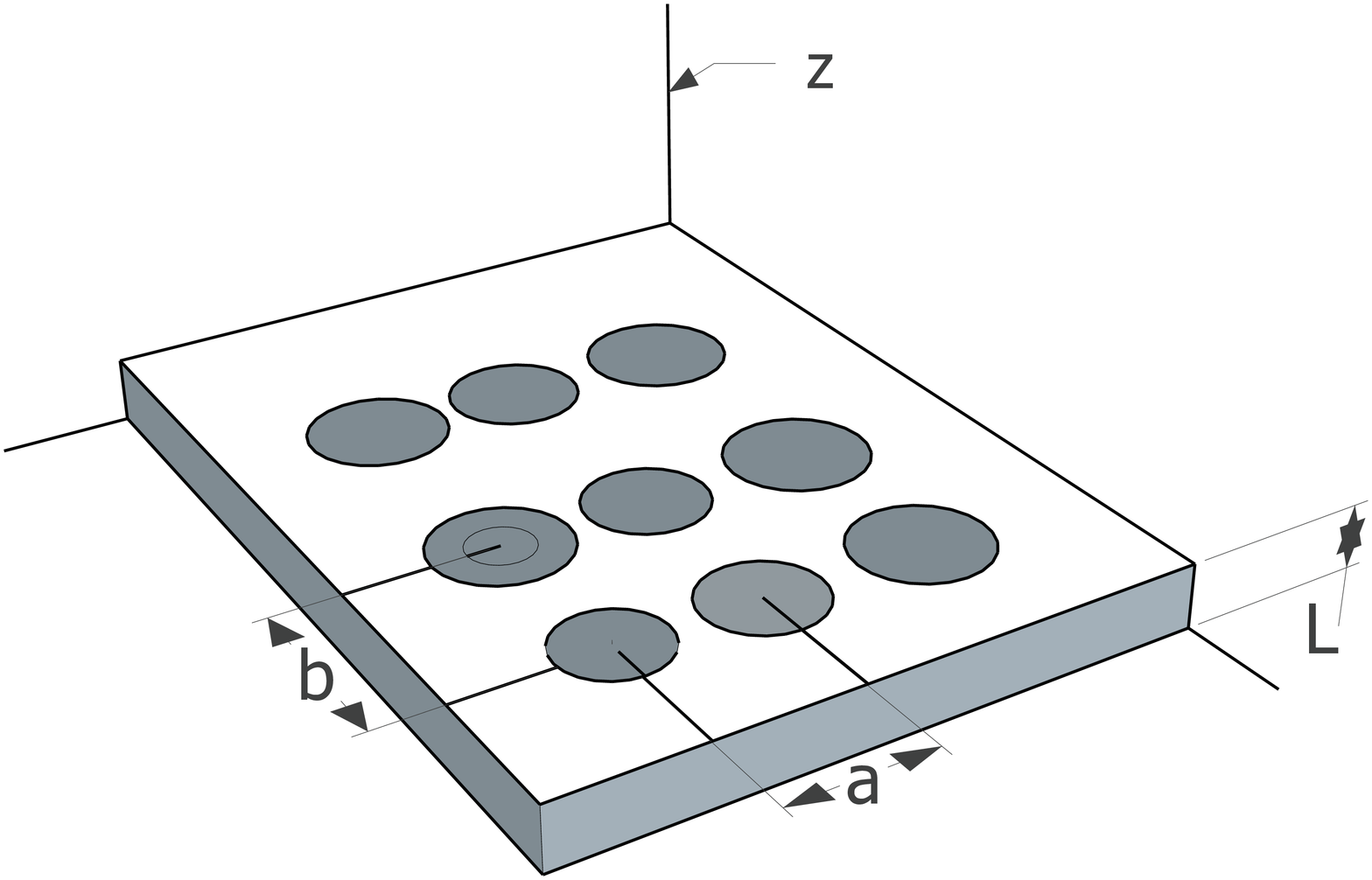}
\caption{Geometry of the problem.}
\label{fig.1}
\end{center}
\end{figure}
 Suppose a  plane wave enters the metal film from the $z<0$ half-space at normal incidence. In order to find the transmission amplitude , we start from a  scalar Helmholtz wave equation
\begin{equation}
\nabla^2\Phi(x,y,z)+k_0^2\varepsilon(x,y)\Phi(x,y,z)=0,
\label{sqeq}
\end{equation}
where $k_0=\omega/c$ is the wave number corresponding to the angular frequency $\omega$ of an incident photon and $\varepsilon(x,y)$ is the two dimensional periodic dielectric permittivity of the system. Eq.(\ref{sqeq}) is valid for s and p-polarized waves and the scalar function $\Phi$ describes the transverse components of an electric or a magnetic fields, respectively.

Mainly following Refs. \cite{FF79,Lag}, we seek the solution of the propagating in the system wave as a product of a fast and a slowly varying, $\phi(x,y,z)$, function on a wave incident $z$ direction, implying that the metal fraction of our system is small.
\begin{equation}
\Phi(x,y,z)=e^{ik_0z}\phi(x,y,z).
\label{slfas}
\end{equation}
Note, that this situation differs from the most of EOT considerations where usually the opposite case is considered.

Substituting Eq.(\ref{slfas}) into Eq.(\ref{sqeq}) and neglecting the second derivative of $\phi$ with respect to $z$ ($|d^2\phi/dz^2|<<2k_0|d\phi/dz|$), one gets
\begin{equation}
i\frac{d\phi}{dz}=\hat{H}(x,y)\phi,
\label{sreqtime}
\end{equation}
where
\begin{equation}
\hat{H}(x,y)=-\frac{1}{2k_0}\nabla_t^2+\frac{k_0}{2}\left(1-\varepsilon(x,y)\right)
\label{tem}
\end{equation}
and $\nabla_t^2\equiv (\partial^2/\partial x^2+\partial^2/\partial y^2)$.

The obvious similarity of Eq. (\ref{sreqtime}) (the spatial coordinate $z$ plays the role of the time) and the time-dependent Scr$\ddot{o}$dinger equation for a particle with mass $k_0$, moving in the two-dimensional potential $V(x,y)=\frac{k_0}{2}\left(1-\varepsilon(x,y)\right)$ may be used as a starting point to evaluate the wave transmission coefficient at $z$.

The solution of Eq.(\ref{sreqtime}) can be represented through the eigenfunctions of Hamiltonian Eq.(\ref{tem})
\begin{equation}
\phi(x,y,z)=\sum_nc_ne^{-iE_nz}\phi_n(x,y)
\label{eig}
\end{equation}
and
\begin{equation}
\hat{H}\phi_n(x,y)=E_n\phi_n(x,y).
\label{eigfun}
\end{equation}

Finally, substitution of Eq.(\ref{eig}) into Eq.(\ref{slfas}) yields the solution of Maxwell equation
\begin{equation}
\Phi(x,y,z)=e^{ik_0z}\sum_nc_ne^{-iE_nz}\phi_n(x,y).
\label{maxsol}
\end{equation}
It follows from Eq. (\ref{maxsol}) that the local transmission amplitude of a central diffracted wave can be defined as
\begin{equation}
t(x,y)=\sum_{E_n<k_0}c_ne^{-iE_nL}\phi_n(x,y),
\label{transamp}
\end{equation}
where $L$ is the system size in the $z$ direction.

Before entering into a more detailed analysis of the local transmission amplitude, let us note that if we ignore the losses and take into account that the metal dielectric constant in the optical region is a real large negative number, then: (i) the potential energy term $V(x,y)=\frac{k_0}{2}\left(1-\varepsilon(x,y)\right)$ in the Hamiltonian Eq.(\ref{tem}) is positive everywhere, (ii) correspondingly, all $E_n$ are also real and non-negative $E_n\geq 0$ and
(ii) exploiting $|d^2\phi/dz^2|<<2k_0|d\phi/dz|$ leads to the condition $E_n<<2k_0$.

The central diffracted wave transmission coefficient that is measured in the experiment can be estimated by using the following expression
\begin{equation}
T=\frac{1}{S}\int dx dy\bigg| t(x,y)\bigg|^2,
\label{trcoeff}
\end{equation}
where $S$ is the area of the system. Substituting Eq.(\ref{transamp}) into  Eq.(\ref{trcoeff}), one has
\begin{equation}
T=\frac{1}{S}\sum_{E_n<k_0}|c_n|^2.
\label{trcoeff2}
\end{equation}
In order to find the coefficients $c_n$ let us consider the Eq.(\ref{maxsol}) for $z=0$
\begin{equation}
\Phi(x,y,z=0)=\sum_nc_n\phi_n(x,y).
\label{zero}
\end{equation}
Next, we assume that the impinging to the system wave has an amplitude $1$ (the region $z<0$).
From the continuity at $z=0$, one has $\Phi(x,y,z=0)=1+r(x,y)$, where $r(x,y)$ is a local reflection coefficient which for the metal without holes is approximately $-1$. Clearly, the existence of the holes will change the value of $r$. However, this change will not affect the further calculations, and for this reason in our further calculations, for the reflection coefficient we assumed some average value $r$ close zero provided that metallic fraction is small. Within this approach, multiplying both sides of Eq.(\ref{zero}) by $\phi_n^*(x,y)$ and integrating over the surface, one has
\begin{equation}
c_n=(1+r)\int dx dy\phi_n^*(x,y).
\label{expcoef}
\end{equation}
Substituting Eq.(\ref{expcoef}) into Eq.(\ref{trcoeff2}), we arrive at the final result for the transmission coefficient
\begin{equation}
T=\frac{|1+r|^2}{S}\sum_{E_n<k_0}\int d\vec\rho d\vec \rho^{\prime}\phi_n^*(\vec\rho)\phi_n(\vec\rho^{\prime}), \label{fincoef}
\end{equation}
where $\vec\rho\equiv (x,y)$ is a two dimensional vector on the $xy$ plane.

This is our main general result. In the following subsections
we analyze its limits for different models.

To close this section let us note that if the dielectric permittivity $\varepsilon(x,y)$ is a periodical function, then the spectrum of Hamiltonian Eq.(\ref{tem}) consists of allowed and forbidden energy bands. As for the transmission coefficient, it depends on the position of incident wavenumber in the transverse energy spectrum.

 \section{Bloch states}
As it follows from Eq.(\ref{fincoef}) the transmission coefficient equals zero provided that $k_0<E_b$, where $E_b$ is the bottom value of first energy band. When $k_0$ lies in the  zone, using  Bloch states transmission coefficient can be rewritten in terms of a quasi-momentum $\vec q$ as
\begin{equation}
T=|1+r|^2I,
\label{tri}
\end{equation}
 where
\begin{equation}
 I=\frac{1}{S}\int_{E(\vec q)<k_0}\frac{d\vec q}{(2\pi)^2}\int d\vec \rho d\vec \rho^{\prime}\phi^*_{\vec q}(\vec\rho)\phi_{\vec q}(\vec\rho^{\prime}).
 \label{invar}
 \end{equation}
Here integration over quasi-momentum $\vec q$ is carried out over the first Brilloin zone $-\pi/a\leq q_x\leq \pi/a, \quad -\pi/b\leq q_y\leq \pi/b$ and $a,b$ are periods of $\varepsilon(x,y)$ in the $x$ and $y$ directions, respectively. According the Bloch theorem the eigenstate $\phi_{\vec q}$ in a periodical potential can be represented in the form
 \begin{equation}
 \phi_{\vec q}(\vec\rho)=e^{i\vec q\vec \rho}u_{\vec q}(\vec \rho),
 \label{eigenfu}
 \end{equation}
 where $u_{\vec q}(\vec \rho)$ is a periodical function satisfying the equation
 \begin{equation}
 \left[-\frac{1}{2k_0}(i\vec q+\vec \nabla)^2+V(\vec \rho)\right]u_{\vec q}(\vec\rho)=E(\vec q)u_{\vec q}(\vec\rho).
 \label{sreq}
 \end{equation}
 For simplicity and as an illustration of the approach, we will carry out further consideration in one-dimensional case. We hope that the results, obtained in this particular case, will enable us to understand, at least in the qualitative level, the transmission of an electromagnetic waves in three-dimensional system in the presence of a transverse waveguide modes tunneling.
\paragraph{Kronig-Penney model}
Suppose that slits are periodically placed in the $x$ axis, which is transverse to the direction of propagation. A cross section of the potential in the $x$ direction can be presented as an array of square potential wells. A metal part will serve as a barrier and characterized by width $b$ and period $a$.  A width of a slit is $a-b$, correspondingly. The metallic dielectric constant described by the Drude model $\varepsilon_m=1-\omega_p^2/\omega^2$ and the height of a barrier is defined as $V_m=k_p^2/2k_0$ ($k_p=\omega_p/c$ and $\omega_p$ is the plasma frequency of a metal). The vacuum part dielectric describes with a $\varepsilon=1$ and with a potential energy $V=0$. For a metal in optical region usually $V_m>k_0$. Because only the energies $E_n<k_0$ give contribution to the transmission coefficient $T$ we will consider the case $E<V_m$ when finding the spectrum of Hamiltonian Eq.(\ref{tem}). The quantum-mechanical problem Eq.(\ref{tem}) is reduced to the well-known Kronig-Penney model \cite{Krpen}.
Let us write $I$, defined by Eq. (\ref{invar}), for one-dimensional case
\begin{equation}
I=\frac{1}{L}\int_{E(q)<k_0}\frac{dq}{2\pi}\int dx dx^{\prime}\phi_q^*(x)\phi_q(x^{\prime}).
\label{onedim}
\end{equation}
Using Bloch theorem $\phi_q(x)=e^{iqx}u_q(x)$, where $u_q(x)$-periodical function one obtains
\begin{equation}
I=\frac{1}{L}\int_{E(q)<k_0}\frac{dq}{2\pi}\sum_{nm}\int_{(n-1)a}^{na}dxe^{-iqx}u_q^*(x)\int_{(m-1)a}^{ma}dxe^{iqx}u_q(x)\nonumber.
\label{parts}
\end{equation}
Changing the variables one finds
\begin{equation}
I=\frac{1}{L_x}\int_{E(q)<k_0}\frac{dq}{2\pi}\sum_ne^{-iqan}\times\sum_me^{iqam}\int_0^adxe^{-iqx}u_q^*(x)\int_0^adxe^{iqx}u_q(x),
\label{first}
\end{equation}
where wave function in the unit cell and in the different regions, is found from Eq.(\ref{sreq})
\begin{eqnarray}
u_{q1}(x)=(A\cos\beta x+B\sin\beta x)e^{-iqx},\quad 0<x<a-b \nonumber \\
u_{q2}(x)=(A\cosh\alpha x+D\sin h\alpha x)e^{-iqx},\quad a-b<x<a
\label{wavefun}
\end{eqnarray}
with $\beta=\sqrt{2k_0E}$ and $\alpha=\sqrt{2k_0(V_m-E)}$.

Substituting $\sum_ne^{-inqa}=2\pi\delta(qa)$ into Eq.(\ref{first}), one obtains
\begin{equation}
I=\frac{1}{a}\left[\int_0^au(x)dx\right]^2
\label{final}
\end{equation}
where $u(x)\equiv u_{q=0}(x)$ is determined by Eq. (\ref{wavefun}). The constants $B, C, D$ can be expressed by $A$ using boundary conditions. $A$ itself can be found from the normalization condition $\int_0^a|u(x)|^2dx=1/N$, where $N=L_x/a$ is the number of unit cells. Using continuity at $x=0$, one gets $C=A$. From the continuity of $du/dx$ at $x=0$, one has $B\beta=D\alpha$. Finally from the continuity of $u(x)$ and $du/dx$ at $x=a-b$ and periodicity $du_1(x)/dx|_{x=a-b}=du_2/dx|_{x=-b}$, one has
\begin{eqnarray}
A\left[\cos\beta(a-b)-\cosh\alpha b\right] +B\left[\sin\beta(a-b)+\frac{\beta}{\alpha}\sinh\alpha b\right]=0\nonumber\\
A\left[\alpha\sinh\alpha b-\beta\sin\beta(a-b)\right] +B\left[\beta\cos\beta(a-b)-\beta\cosh\alpha b\right]=0.
\label{bound}
\end{eqnarray}
Equalizing the determinant of $2\times 2$ homogeneous equation to 0 one gets the dispersion relation for $q=0$
\begin{equation}
1=\frac{\alpha^2-\beta^2}{2\alpha\beta}\sinh\alpha b\sin\beta(a-b)+\cosh\alpha b\cos\beta(a-b).
\label{disp}
\end{equation}

Expressing now coefficient $B$ through $A$ according to Eq. (\ref{bound}) and taking elementary integrals we finally arrive at the expression for $I$:

\begin{eqnarray}
I=\frac{8}{\alpha\beta a}\frac{\bigg [{\alpha\tan{\frac{\beta(a-b)}{2}}+\beta \tanh{(\alpha b/2)}\cosh{\alpha a}}\bigg]^2}{{{2\alpha}}\tan\frac{\beta(a-b)}{2}+{2\beta}\tanh\frac{\alpha b}{2}\cosh 2\alpha a+{a\alpha\beta}+\alpha\beta\tan^2\frac{\beta(a-b)}{2}
\bigg({a-b}-{\beta^2b}{\alpha^2}
\bigg)}\label{FA}.
\end{eqnarray}
Note that above expression has been derived using the first relation of Eq. (\ref{bound}). However, using dispersion relation (\ref{disp}), one can show easily that the second relations leads the same result.
In figure 2 we present the function $I$ versus $b/a$ for $k_p=4.6\times10^{-2}nm^{-1}$ (silver $Ag$) and $k_p=1.84\times 10^{-2}nm^{-1}$ (potassium $K$) (in both cases the wave number of an incident photon $k_0=10^{-2}nm^{-1}$). We have checked that presented curves are unaffected by the change of $k_0$ in the visible region $0.62\times 10^{-2}<k_0<1.57\times 10^{-2}nm^{-1}$. This means that for a given period $a$ and $k_p$, all the curves can be scaled into a single curve. The physical reason of the $k_0$ independent
of the results is in the structure of the dispersion relation (\ref{disp}). The latter, in the frame of the adopted approach, can be described by a mean of the unique combination $k_0E$.

\begin{figure}
\vspace{-3cm}
 \begin{center}
\includegraphics[width=8.0cm]{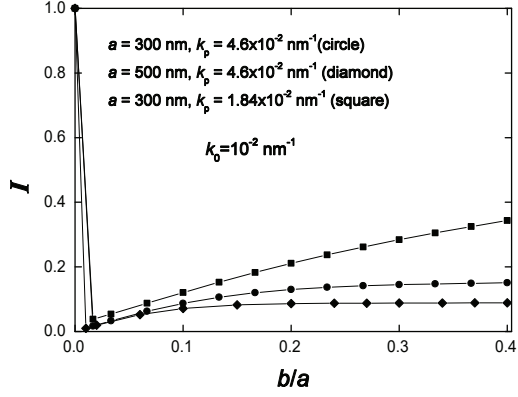}
\caption{Plot of the function $I(b/a)$ on a metal fraction for two values of $k_p$ (silver and potassium). The wave number of an incident photon $k_0=10^{-2}nm^{-1}$.}
\label{fig.2}
\end{center}
\end{figure}
We now turn to the numerical calculations of the $I$ dependence on $b/a$ for some typical values of $a=100-500nm$ and $b=10-500nm$ in visible region $0.62\times 10^{-2}<k_0<1.57\times 10^{-2}nm^{-1}$. The results are plotted in Fig.2.
As follows from Fig. 2, even a small
amount of metal is enough to essentially reduce $I$ from its maximal value 1 to almost 0 around an origin. This jump happens on a short scale ($b/a \approx 0.01$), where three curves reached their minima. The minimum can be explained by diffraction, which spreads and reduces the initial light intensity randomly across the entire system. For silver, (circle and diamond symbols) $I$ increases very slowly with increasing a fraction of metal at the beginning and remains almost flat with further increasing of $b/a$. However, for potassium with relatively small metal barrier height and with large tunneling rate across a barrier, the central diffracted wave transmission coefficient (square  symbol) increases with further increasing of a metal fraction (see Fig. 2). This leads to some focusing effect 
in a central diffraction direction.

Note, that the transmission coefficient $T$, Eq. (\ref{tri}), includes also the multiplier $|1+r|^2$ which in contrast to the $I$, decreases with increasing $b/a$. However, the rate of decreasing of $|1+r|^2$ is much slower than the rate of increasing of $I$, and as a result the central diffracted wave transmission coefficient is increasing with $b/a$.
Such an unusual behavior is caused by periodicity and Bloch states. It is easy to get convinced that in the case of isolated slits, where the wave functions overlap is negligible and where Bloch states cannot exit, the $T$ behaves totally differently.
\section{Isolated slits}
For large plasmonic wavenumber $k_p$ band's width becomes very narrow. The reason for it is that $k_p$ determines the barriers height and large $k_p$ suppresses tunneling through the barrier. In this case the transversal wave functions become less and less extended in space and more localized within a holes with negligible overlap.
In this limiting case one can use the infinite potential well approximation to evaluate the transmission coefficient (\ref{fincoef}).
Writing the wave functions in the form \cite{box}
\begin{equation}
\phi_n(x)=\sqrt{\frac{2}{a-b}}\sin\frac{n\pi x}{a-b}
\label{box}
\end{equation}
and substituting Eq.(\ref{box}) into Eq.(\ref{fincoef}), we find
\begin{equation}
T_{is}=\frac{8|1+r|^2}{\pi^2}\frac{a-b}{a}
\label{isol}
\end{equation}
We arrived at the above expression summing over all the independent slit contribution and restricted ourselves by terms $n=1$ while calculating sum in Eq.(\ref{fincoef}). The contributions of terms with $n>1$ become irrelevant because $k_0< E_n$ and therefore only the first band give contribution to the
transmission.

Comparing Eq.(\ref{isol}) with the maximal value, one has $T_{is}/T_{max}\sim \frac{a-b}{a}$. As expected, for isolated holes the transmission coefficient becomes size dependent, that is $T_{is}/T_{max}$ is proportional to the fraction of the vacuum part in the system. In this sense, the isolated holes system reveals a usual dependence of transmission coefficient on metal fraction. We expect to find the same ratio be valid also in the case of two dimensional hole array.


\section{ Disordered hole arrays}
In this case it is convenient to represent the transmission coefficient, Eq.(\ref{fincoef}), in the form
\begin{equation}
T=\frac{|1+r|^2}{S}\int_0^{k_0}d E<\sum_n\delta(E-E_n)|\phi_n(0)|^2>
\label{dis}
\end{equation}
where $<...>$ means averaging over random positions of holes and $\phi_n(\vec q)$ is the Fourier transform of $\phi_n(\vec r)$ satisfying Schr$\ddot{o}$dinger equation with random potential
\begin{equation}
\bigg[-\frac{1}{2k_0}\nabla_t^2+V(\vec r)\bigg]\phi_n(\vec r)=E_n\phi_n(\vec r)
\label{ranshred}
\end{equation}
where $V(\vec r)$ is assumed to be Gaussian distributed random function with a correlator $B$
\begin{equation}
<(V(\vec r)-\overline{V})(V(\vec r')-\overline{V})>=B(|\vec r-\vec r'|)
\label{corfun}
\end{equation}
where $\overline{V}=\frac{1}{S}\int d\vec r V(\vec r)=k_0(1-\varepsilon_m)(1-f_v)/2$ and $f_v$ is the fraction of the vacuum part (in 1D periodic system, discussed in the previous subsection $1-f_v=b/a$).

We now turn to the calculation of the transmission coefficient, Eq.(\ref{dis}). In order to carry out averaging over the randomness, it is
convenient the latter quantity to express through the average Green’s function
\begin{equation}
T=|1+r|^2\int_0^{k_0}\frac{dE}{\pi}<-ImG_E(q=0)>
\label{imgren}
\end{equation}
with $G_E=[E-H+i\delta]^{-1}$.

The averaged Green's function can be represented in the form \cite{theo}
\begin{equation}
<G_E(\vec q)>=\frac{1}{E+\overline{V}-\Sigma}
\label{avgrfun}
\end{equation}
where $\Sigma=\sum_{n\geq 2}\Sigma_n$ is the self-energy constituting contributions of irreducible parts of different order. In further we will restrict ourselves by the first term in the sum
\begin{equation}
\Sigma_2(\vec q)=\int\frac{d\vec k}{(2\pi)^2}B(|\vec q-\vec k|)G_0(k)
\label{bornap}
\end{equation}
where $G_0(q)=[E-q^2/2k_0+i\delta]^{-1}$ is the bare Green's function.

The explicit form of correlation function $B(q)$ is needed to obtain a closed analytical expression for T, Eq. (\ref{imgren}). Particularly,  in the limit of very small size $h\to 0$ holes, $B(q)$ can be substituted by $B(q=0)=B_0\sim f_vk_0^2h^2(1-\varepsilon_m)^2$. Evaluating the integrals Eqs.(\ref{imgren}) and (\ref{bornap}) we find
\begin{equation}
T_d=\frac{|1+r|^2}{\pi}\left[\arctan\frac{2(k_0+\overline{V})}{k_0B_0}-\arctan\frac{2\overline{V}}{k_0B_0}\right]
\label{distrans}
\end{equation}
When obtaining Eq.(\ref{distrans}) we neglect $Re\Sigma$ relative to $\overline{V}$.
Expanding arctan functions in the limit $B_0\to 0$, for the transmission coefficient in the disordered case, we finally obtain
\begin{equation}
T_d=\frac{|1+r|^2}{2\pi}\frac{k^2_0B_0}{\overline{V}(k_0+\overline{V})}
\label{disfin}
\end{equation}
Comparison of the Eqs.(\ref{isol}) and (\ref{disfin}) shows important differences
 between two cases, in spite of and due to the presence of the formal factor ${|1+r(\omega)|^2}$ (see. Ref. \cite{disor}).
In disordered case, Eq. (\ref{disfin}), in contrary to periodical case, $r(\omega)$ has no peculiarities and is a smooth function of $\omega$.
This means that the randomness destroy the resonant spectral shape and lead to its broadening \cite{disor,AuB09}, assuming that the second multiplier in Eq. (\ref{disfin}) is smoothly varying function of the frequency. This is true for almost all accepted metallic models and can be seen using the explicit form of the dielectric constant $1-\varepsilon_m=\omega^2_p/\omega^2$, $B_0$ and $\overline{V}$.

By comparing with isolated case contribution, Eq.(\ref{isol}) and assuming that $\overline{V}\gg k_0$ and $B_0\to 0$, one finds
\begin{equation}
\frac{T_d}{T_{is}}\sim k_0^2h^2
\label{ratio}
\end{equation}
It follows from Eq.(\ref{ratio}) that disordered case the transmission coefficient $T_d$ much smaller than $T_{is}$ provided that $k_0h\ll 1$. Two coefficients become of the same order when $k_0h\sim 1$.
\section{Conclusion and Discussion}
We have discussed the problem of light transport through a perforated metal, taking into account the transverse waveguide modes tunneling in two-dimensional plane perpendicular to traveling direction of light.
Periodic, isolated and disordered holes systems are analyzed in details. Analytical expressions are derived for all different regimes. The one dimensional case of periodic holes, due to its simplicity, is investigated in detail. In the dilute metal regime, when metal fraction is small, our numerical study of the transmission coefficient of central diffracted wave indicates the existence of a minimum which is completely independent of an incident wavelength. The transmission coefficient of central diffracted wave increases when metal fraction of the system is increasing. The main contribution to the transmission coefficient is connected with extended states that close to the center $q=0$ of the Brillouin zone. This means that in order to observe the above mentioned peculiarities in perforated systems, it is enough that the system exhibits long or quasi-long-range structural order in $xoy$ plain (see also Ref.\cite{longorder}).

In our discussion we take into account the influence of a transverse tunneling between different holes on the transmission coefficient $T$.  As a result, $T$ does not depend on the system thickness in $z$ direction(we ignore the imaginary part of $\varepsilon$). Obviously, exponential decaying of $T$ with thickness will arise on if one takes into account the losses.

 In our consideration we substitute a local reflection coefficient by an average value. This assumption seems more
 relevant in the random hole arrays case. However, even in the periodical case one can imagine $r(\omega)$ as a quantity that correctly takes into account periodical hole arrays similar to periodical gratings \cite{plasmon}.
Note that all plasmonic effects are included into $r(\omega)$. Particularly on impinging of p-polarized light plasmon is generated on the perforated surface. Reflection coefficient close to the plasmonic resonance becomes minimal \cite{plasmon} leading to the maximal value of transmission coefficient. Plasmonic resonance takes place when plasmon wavenumber coincides with one of the photonic crystal reciprocal lattice periods,see for example, \cite{GG14}. Note that $r(\omega)$ depends on the geometry of the perforated surface and can detect resonance effects associated with the geometry of holes \cite{geom}.

\section{ Acknowledgments.}
V.G. and E.C. acknowledge partial support by FEDER and the Spanish DGI under Project no. FIS2010-16430. Zh.G. is grateful to A.Hakhoumian, A.Allahverdian,D.Karakhanian, Kh.Nerkararian and T.Shakhbazian for helpful discussions.

\section*{References}

\bibliography{mybibfile}

\end{document}